\documentclass[12pt]{article}

\usepackage{algorithm}				
\usepackage[noend]{algpseudocode}	
\usepackage{amsmath}				
\usepackage{amssymb}				
\usepackage{amsthm}				
\usepackage{caption}
\usepackage{color}
\usepackage[margin=1in]{geometry}
\usepackage[hidelinks]{hyperref}
\usepackage{titling}

\usepackage{graphicx}

\usepackage{tikz}
\usetikzlibrary{automata,positioning}

\theoremstyle{plain}				

\theoremstyle{definition}			

\theoremstyle{remark}			

\thanksmarkseries{alph}

\DeclareRobustCommand{\acro}[1]{{\small\textsc{#1}}}
\DeclareRobustCommand{\cs}[1]{\texttt{#1}}

\def\plus{\raisebox{.7ex}{$_{+}$}}	
\def\Cplusplus{C\plusplus}
\def\PDF{\acro{PDF}}

\def\plusplus{\raisebox{.7ex}{$_{++}$}}
\def\texttub{\textsl}
\def\TikZ{Ti\/{\em k}Z}
\def\TUB{\texttub{TUGboat}}

\title{Illustrating Finite Automata with Grail\plus\ and \TikZ}
\author{
	Alastair May \thanks{Department of Computer Science, St.\ Francis Xavier University, Antigonish, Nova Scotia, Canada. Email: \href{mailto:x2016owd@stfx.ca}{\texttt{x2016owd@stfx.ca}}.}
	\and
	Taylor J. Smith \thanks{Department of Computer Science, St.\ Francis Xavier University, Antigonish, Nova Scotia, Canada. Email: \href{mailto:tjsmith@stfx.ca}{\texttt{tjsmith@stfx.ca}}.}
}
\date{\today}

\begin{document}


\maketitle

\begin{abstract}
In this article, we discuss a new software tool that interacts with Grail\plus, a library of automata-theoretic command-line utilities. Our software, the Grail\plus\ Visualizer, takes the textual representation of a finite automaton produced by Grail\plus\ and generates \TikZ\ code to illustrate the finite automaton, with automatic layout of states and transitions. In addition to giving an overview of the basics of automata theory and Grail\plus, we discuss how the Grail\plus\ Visualizer works in detail and suggest avenues for future work.

\medskip


\medskip

\noindent\textit{MSC2020 classes:} 68-04 (primary); 68Q45 (secondary).
\end{abstract}


\section{Introduction}

Grail\plus\ is a \Cplusplus\ library of command-line utilities that performs symbolic manipulation of various models of finite automata, regular expressions, and finite languages. Each utility, called a \emph{filter} in Grail\plus\ terminology, can either handle input directly or be piped together to create a chain of filters. Grail\plus\ consists of nearly one hundred filters that can compute common operations or procedures that a theoretical computer scientist might want to perform; among many other tasks, these filters can enumerate the elements of a language (\texttt{fmenum}), convert a finite automaton to an equivalent regular expression (\texttt{fmtore}) and vice versa (\texttt{retofm}), and minimize (\texttt{fmmin}) or determinize (\texttt{fmdeterm}) a finite automaton. These predefined filters, together with the ability to chain filters together, allow users to perform thousands of formal language and automata-theoretic tasks.

The original environment, Grail, was developed by Darrell Raymond at the University of Waterloo and Derick Wood at the University of Western Ontario~\cite{RaymondWood1994Grail}. Following major changes instituted in version 3.0, the name of the environment changed to Grail\plus\ and coordination of the development work was taken over by Sheng Yu, again at the University of Western Ontario. Presently, Grail\plus\ is being developed and maintained by Cezar C\^{a}mpeanu at the University of Prince Edward Island, and the current stable version is 3.4.5~\cite{GrailWebsite}.


\section{Finite Automata in Grail\plus}

Since Grail\plus\ runs in a command-line interface, all input and output is plain text. For models that can be represented naturally in text, like regular expressions, Grail\plus\ follows the conventional notation in theoretical computer science; for instance, the union of regular expressions \texttt{a} and \texttt{b} is \texttt{a + b}, and the concatenation of these regular expressions is \texttt{ab}. For non-textual models like finite automata, however, Grail\plus\ defines its own convention.

\begin{figure}
\centering
\begin{tikzpicture}[node distance=2cm]
\node[state, initial] (q0) {$q_{0}$};
\node[state, accepting, right of=q0] (q1) {$q_{1}$};
\path[->] (q0) edge[loop above] node[above] {\texttt{a}} (q0);
\path[->] (q0) edge[bend left] node[above] {\texttt{b}} (q1);
\path[->] (q1) edge[bend left] node[below] {\texttt{b}} (q0);
\path[->] (q1) edge[loop above] node[above] {\texttt{a}} (q1);
\end{tikzpicture}
\caption{An example of a finite automaton.}
\label{fig:finiteautomaton}
\end{figure}
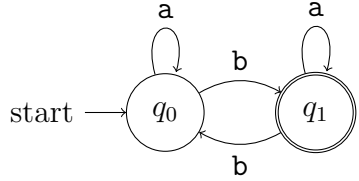

As an example, consider the finite automaton depicted in Figure~\ref{fig:finiteautomaton}. This finite automaton takes a \emph{word} (or \emph{string}) consisting of \texttt{a}s and \texttt{b}s as its input. It has two \emph{states}, labelled $q_{0}$ and $q_{1}$, and \emph{transitions} on these states labelled by \texttt{a} and \texttt{b}. The contents of the word determine which transition the finite automaton follows at any given step of its computation; for example, if the finite automaton reads the symbol \texttt{b} in the word while it is in state $q_{0}$, then it will transition to state $q_{1}$ before reading the next symbol. The state $q_{0}$ is the \emph{initial state}, or the state where the computation of the finite automaton begins. The state $q_{1}$ is a \emph{final} or \emph{accepting state}; if, after reading all symbols in the word, the finite automaton finds itself in state $q_{1}$, then it accepts that word. The set of all words accepted by a finite automaton is the \emph{language} of that finite automaton.

In Grail\plus, finite automata are represented as lists of \emph{instructions}. Each individual instruction consists of three pieces of information: a source state, a label, and a sink state. Additionally, there are two special \emph{pseudo-instructions} to indicate which states are initial and which states are final. Returning to Figure~\ref{fig:finiteautomaton}, this finite automaton would be represented in Grail\plus\ as the following list:
\begin{verbatim}
(START) |- 0
0 a 0
0 b 1
1 a 1
1 b 0
1 -| (FINAL)
\end{verbatim}
There is no particular ordering to the instructions in the lists produced by Grail\plus\ as output, and no ordering is enforced when giving a list to Grail\plus\ as input. Additionally, while the example given here has one initial state and one final state, a finite automaton may contain multiple initial or final states.


\section{Typesetting Finite Automata}

For authors who wish to include illustrations of finite automata in their documents, the most straightforward way to do so---apart from importing an external image file---is to use \TikZ\ and the \acro{PGF} package~\cite{PGFTikZPackage} to create the illustration. \TikZ\ includes an automata drawing library with special shapes and styles specific to finite automata (see Chapter~43 of the \TikZ\ \& \acro{PGF} manual~\cite{Tantau2023TikZPGFManual}; see also the article in \TUB~44:1 by Igor Borja~\cite{Borja2023TikZAutomata}). An example of \TikZ\ code using the automata drawing library is given in Figure~\ref{fig:finiteautomatontikz}.

\begin{figure}
\begin{verbatim}
\begin{tikzpicture}[node distance=2cm]
    \node[state, initial] (q0) {$q_{0}$};
    \node[state, accepting, right of=q0] (q1) {$q_{1}$};
    \path[->] (q0) edge[loop above] node[above] {\texttt{a}} (q0);
    \path[->] (q0) edge[bend left] node[above] {\texttt{b}} (q1);
    \path[->] (q1) edge[bend left] node[below] {\texttt{b}} (q0);
    \path[->] (q1) edge[loop above] node[above] {\texttt{a}} (q1);
\end{tikzpicture}
\end{verbatim}
\caption{The (human-written) \TikZ\ code producing the finite automaton in Figure~\ref{fig:finiteautomaton}.}
\label{fig:finiteautomatontikz}
\end{figure}

Creating illustrations from scratch using \TikZ\ puts the decision-making entirely in the user's hands, allowing for ample customization in layout, style, and other aspects. At the same time, creating illustrations from scratch using \TikZ\ puts the decision-making \emph{entirely} in the user's hands, leading to potential pitfalls. For particularly large or complicated finite automata, laying out states and transitions in an aesthetically pleasing way can become very difficult. Thus, using external software to assist in constructing and laying out the finite automaton can make the illustration process easier.

Andrew Mertz, William Slough, and Nancy Van Cleave wrote in \TUB~35:2 about methods of illustrating computer science concepts using \LaTeX\ packages~\cite{Mertz2014TypesettingFiguresCS}. In particular, in Section~7 of their article, the authors discuss typesetting automata using \acro{JFLAP}~\cite{RodgerFinley2006JFLAP}, which is a Java software package for manipulating finite automata and formal languages. In contrast to Grail\plus, \acro{JFLAP} uses a graphical user interface, and it is capable of exporting illustrations of finite automata to various image formats. The \texttt{jflap2tikz} package~\cite{JFLAP2TikZPackage} additionally allows users to convert a finite automaton produced by \acro{JFLAP} to \TikZ\ code that can be included in a \LaTeX\ document.


\section{Motivation}

While \acro{JFLAP} is a popular software package, and while it is capable of handling more theoretical models of computation than Grail\plus\ currently handles---namely grammars, pushdown automata, and Turing machines---there are reasons why users may still prefer to work with Grail\plus. For one, Grail\plus\ can be run on any computer that is capable of compiling \Cplusplus\ code, and the suite can be customized and extended by anyone who is capable of writing \Cplusplus\ code. Subjectively, users may find the text-based command-line interface of Grail\plus\ to be faster or easier to use than a point-and-click graphical user interface. Lastly, while software like \acro{JFLAP} emphasizes pedagogy and learning about formal languages and automata theory, researchers and practitioners may value Grail\plus\ for its efficient implementation and wider array of features specific to finite automata. (Together with these reasons, Canadians may uniquely value Grail\plus\ for being made-in-Canada software.)

What Grail\plus\ lacks, however, is a way to render its textual output in a more human-friendly form. It can be extremely difficult, especially with finite automata having many states or many transitions, for a user to parse the textual output and to gain an understanding of the structure of the finite automaton without representing it visually. Given that \TikZ\ has a built-in automata drawing library, and that both the output from Grail\plus\ and the \TikZ\ code to produce an illustration of a finite automaton follow a fixed, pre-specified format, \TikZ\ is a natural candidate for automatically laying out illustrations of finite automata produced by Grail\plus.


\section{The Grail\plus\ Visualizer}

Our software tool, the Grail\plus\ Visualizer~\cite{GrailVisualizerRepository}, offers a more human-friendly---and beginner-friendly---way to construct, manipulate, and display automata. Working alongside Grail\plus, our visualizer transforms the textual output of Grail\plus\ into \TikZ\ code that can be either typeset and displayed on its own or inserted into an existing \LaTeX\ document.

The visualizer software is written in Bash and can be run directly from a command-line interface, just like Grail\plus\ itself. The software can therefore act as the final link in a chain of Grail\plus\ filters, providing an immediate visual indication of the result.


\subsection{How the Visualizer Works}\label{subsec:how}

The Grail\plus\ Visualizer takes as input the textual representation of a finite automaton produced by Grail\plus\ (i.e., the list of instructions) and parses the text to extract state labels, which are stored in a list. At the same time, state type labels are stored in an auxiliary list to distinguish whether a particular state is initial, final, or both. Transitions are also parsed and stored in three lists: one containing source state labels, one containing transition labels, and one containing sink state labels.

The visualizer then begins to lay out states and transitions. Each state is placed at coordinate $(x,y)$ on a square grid according to the following procedure:
\begin{enumerate}
\item Assign $x$-coordinates to states in the order they are read from the input; that is, the first state read from input is assigned $x = 0$, the second state is assigned $x = 1$, and so on.
\item Initialize an all-zero array $A$ of size $|Q|$, where $|Q|$ is the number of unique states identified in the input processing step. Indices of $A$ correspond to $x$-coordinates of states; for example, $A[0]$ corresponds to the state having $x$-coordinate $0$. Entries in $A$ correspond to $y$-coordinates of states. At this point, each state has an initial $y$-coordinate of $0$.
\item For each pair of states $p$ and $q$ connected by a transition in the finite automaton, where $p$ has $x$-coordinate $i$, $q$ has $x$-coordinate $j$, and assuming $i < j$ without loss of generality:
	\begin{enumerate}
	\item Denote by $m$ the maximum entry in the subarray $A[i..j]$.
	\item Increment $m$ by $1$.
	\item Compare the value $m$ to the entries $A[i]$ (i.e., the $y$-coordinate of state $p$) and $A[j]$ (i.e., the $y$-coordinate of state $q$). The largest of these three values will be the new $y$-coordinate of both states $p$ and $q$.
	\end{enumerate}
\end{enumerate}
These coordinates are used to generate a list of \TikZ\ \cs{node}s, which is written to the output file. The label of each node corresponds to the state label stored during the input processing step. If a state is distinguished as initial or final, then the appropriate option (\texttt{initial} or \texttt{accepting}, respectively) is added to the corresponding node.

Next, where a transition exists from a source state $p$ to a sink state $q$, the visualizer writes a \TikZ\ \cs{path} to the output file producing a directed edge from $p$ to $q$. This path is labelled by the transition label stored during the earlier input processing step. Some special cases are also handled during this step:
\begin{itemize}
\item Where there exists a transition from a state $p$ to itself, the visualizer adds the option \texttt{loop above} to the corresponding path.
\item Where there exist multiple transitions between a source state $p$ and a sink state $q$, those multiple transitions are consolidated into a single path. The labels of the multiple transitions are obtained via pattern-matching existing lines of the output file and stored in a temporary variable. The existing paths are deleted, and a new path labelled by the stored transition labels is written to the output file.
\end{itemize}

At this stage, the \TikZ\ code is complete, and the output file is ready for compilation. The file is typeset by \PDF\LaTeX\ and can be opened by a \PDF\ viewer immediately, or the \TikZ\ code may be copied into another \LaTeX\ document.


\subsection{An Example}\label{subsec:example}

To demonstrate some of the capabilities of the Grail\plus\ Visualizer, consider the following list of instructions produced by Grail\plus:
\begin{verbatim}
(START) |- 0
(START) |- 6
(START) |- 3
0 a 1
1 c 3
1 d 5
6 b 7
7 c 9
7 d 11
3 -| (FINAL)
5 -| (FINAL)
9 -| (FINAL)
11 -| (FINAL)
\end{verbatim}
In this list of instructions, we have eight states unordered and numbered non-consecutively (e.g., there is a state \texttt{1} and a state \texttt{3}, but no state \texttt{2}). Of these states, three are initial states and four are final states, and state \texttt{3} is both initial and final. There is also a total of six transitions.

The Grail\plus\ Visualizer begins by parsing the list of instructions and extracting the state labels from these instructions. 
From this input, the visualizer produces two preliminary lists: one of state labels and one of state types. At this stage, the state labels list may contain duplicates. For instance, state \texttt{3} appears three times in the preliminary state label list, corresponding to its appearances in lines 3, 5, and 10 of the input list of instructions.

The visualizer also extracts transition data and produces an additional three lists: \\
\begin{tabular}{c c c c c c}
\texttt{0} & \texttt{1} & \texttt{1} & \texttt{6} & \texttt{7} & \texttt{7} \\
\texttt{a} & \texttt{c} & \texttt{d} & \texttt{b} & \texttt{c} & \texttt{d} \\
\texttt{1} & \texttt{3} & \texttt{5} & \texttt{7} & \texttt{9} & \texttt{11}
\end{tabular} \\
From top to bottom, these three lists indicate the transition source states, the transition labels, and the transition sink states.

Duplicate state labels are then removed from the preliminary lists created earlier. In this example, the unique state labels are identified in the order \texttt{0}, \texttt{6}, \texttt{3}, \texttt{1}, \texttt{5}, \texttt{7}, \texttt{9}, and \texttt{11}, 
and the lists produced by the Grail\plus\ Visualizer are as follows: \\
\begin{tabular}{c c c c c c c c}
\texttt{0} & \texttt{6} & \texttt{3} & \texttt{1} & \texttt{5} & \texttt{7} & \texttt{9} & \texttt{11} \\
S & S & B & -- & F & -- & F & --
\end{tabular} \\
The top list contains state labels, while the bottom list contains state types. States may have one of four types: ``S'' denotes an initial state, ``F'' denotes a final state, ``B'' denotes a state that is both initial and final, and ``--'' denotes an undistinguished state.

Next, state positions are calculated according to the procedure outlined in Section~\ref{subsec:how}. This procedure assigns the following $y$-coordinates to states: \\
\begin{tabular}{c c c c c c c c}
\texttt{0} & \texttt{6} & \texttt{3} & \texttt{1} & \texttt{5} & \texttt{7} & \texttt{9} & \texttt{11} \\
1 & 4 & 0 & 3 & 0 & 6 & 0 & 6
\end{tabular} \\
From top to bottom, these lists indicate the state label and that state's $y$-coordinate. Recall that the state's $x$-coordinate is its index in the list, so state \texttt{0} is at position $(0,1)$, state \texttt{6} is at position $(1,4)$, and so on.

At this point, the various lists created by the visualizer are then used to produce the \TikZ\ code corresponding to the automaton. The final \TikZ\ code produced by the visualizer is shown in Figure~\ref{fig:graillatex}, and typesetting the code produces the automaton shown in Figure~\ref{fig:graillatexoutput}.


\section{Conclusions and Future Work}

In this article, we introduced the Grail\plus\ Visualizer, explained how the visualizer interacts with Grail\plus\ to produce a typeset illustration of a finite automaton, and worked through an example demonstrating how the visualizer works. We hope that this software tool stimulates a greater interest in both automata theory and the Grail\plus\ library.

There remain some areas for improvement in the visualizer software. For instance, the layout procedure can be optimized to place states sharing many transitions closer to one another or to avoid large numbers of crossing transitions. For certain finite automata, a more optimized layout procedure might use a ``system of springs" technique inspired by that of Tutte~\cite{Tutte1963DrawAGraph}. Another desirable feature might allow the user to specify some degree of customization for the output; say, in setting the colours of states, the minimum distance between states, or the exact positions of certain states. Lastly, implementing the Grail\plus\ Visualizer as a \LaTeX\ package rather than as a separate piece of software would greatly simplify the workflow for users. This could allow a user to write a command like \verb|\grailautomaton{instruction_list}| that would generate and include \TikZ\ code within any \LaTeX\ document at typesetting time.


\section*{Acknowledgements}

The work done by the first author of this article was supported by an Alley Heaps undergraduate research internship provided by St.\ Francis Xavier University.


\bibliographystyle{plain}
\bibliography{references}


\newpage

\begin{figure}
\begin{verbatim}
\begin{tikzpicture}[node distance=2cm]

\node[state,initial] (0) at (0,1) {$0$};
\node[state,initial] (6) at (1,4) {$6$};
\node[state,initial,accepting] (3) at (2,0) {$3$};
\node[state] (1) at (3,3) {$1$};
\node[state,accepting] (5) at (4,0) {$5$};
\node[state] (7) at (5,6) {$7$};
\node[state,accepting] (9) at (6,0) {$9$};
\node[state] (11) at (7,6) {$11$};
\path[->] (0) edge[] node[align=center, anchor=center, above, sloped] {a} (1);
\path[->] (1) edge[] node[align=center, anchor=center, above, sloped] {c} (3);
\path[->] (1) edge[] node[align=center, anchor=center, above, sloped] {d} (5);
\path[->] (6) edge[] node[align=center, anchor=center, above, sloped] {b} (7);
\path[->] (7) edge[] node[align=center, anchor=center, above, sloped] {c} (9);
\path[->] (7) edge[] node[align=center, anchor=center, above, sloped] {d} (11);

\end{tikzpicture}
\end{verbatim}
\caption{\TikZ\ code produced by the Grail\plus\ Visualizer from the example input of Section~\ref{subsec:example}.}
\label{fig:graillatex}
\end{figure}

\begin{figure}
\centering
\begin{tikzpicture}[node distance=2cm]

\node[state,initial] (0) at (0,1) {$0$};
\node[state,initial] (6) at (1,4) {$6$};
\node[state,initial,accepting] (3) at (2,0) {$3$};
\node[state] (1) at (3,3) {$1$};
\node[state,accepting] (5) at (4,0) {$5$};
\node[state] (7) at (5,6) {$7$};
\node[state,accepting] (9) at (6,0) {$9$};
\node[state] (11) at (7,6) {$11$};
\path[->] (0) edge[] node[align=center, anchor=center, above, sloped] {a} (1);
\path[->] (1) edge[] node[align=center, anchor=center, above, sloped] {c} (3);
\path[->] (1) edge[] node[align=center, anchor=center, above, sloped] {d} (5);
\path[->] (6) edge[] node[align=center, anchor=center, above, sloped] {b} (7);
\path[->] (7) edge[] node[align=center, anchor=center, above, sloped] {c} (9);
\path[->] (7) edge[] node[align=center, anchor=center, above, sloped] {d} (11);

\end{tikzpicture}
\caption{The illustration produced by compiling the \TikZ\ code in Figure~\ref{fig:graillatex}.}
\label{fig:graillatexoutput}
\end{figure}
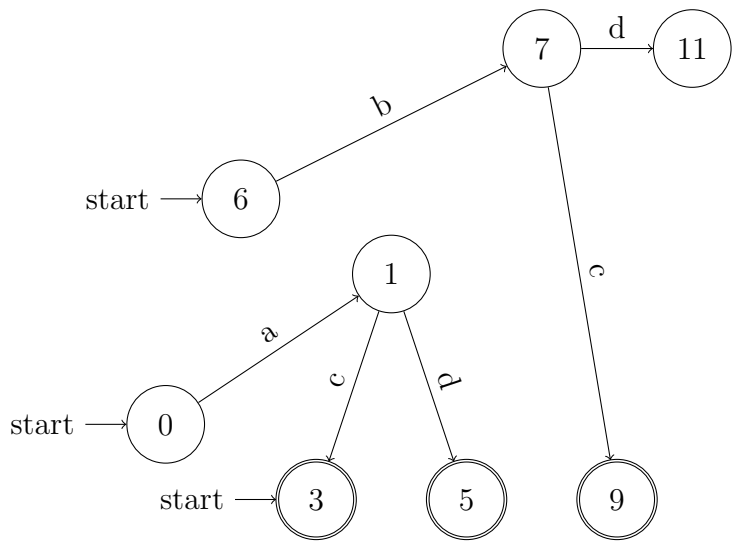


\end{document}